\documentclass[%
 reprint,
 amsmath,amssymb,
 aps,
]{revtex4-2}
\usepackage{braket}
\usepackage{longtable}
\usepackage{physics}
\usepackage{enumitem}
\usepackage{booktabs}
\setlist[enumerate]{itemsep= 0mm}
\setlist[itemize]{itemsep=-0.5 mm}
\usepackage{graphicx}
\usepackage{dcolumn}
\usepackage{bm}

\begin{document}

\preprint{APS/123-QED}

\title{A MATLAB based modeling and simulation package for DPS-QKD}

\author{Anuj Sethia}

\author{Anindita Banerjee}%
\affiliation{QuNu Labs Pvt Ltd., MG Road, Bangalore, India}



\date{\today}

\begin{abstract}
Quantum key distribution (QKD) is an ingenious technology utilizing quantum information science for  provable secure communication. However, owing  to the technological limitations and device non-idealities it is  important to analyze the system performance critically and carefully define the  implementation security. With an acceleration in the commercial adoption of QKD, a simulation toolkit is requisite to evaluate the functional architecture of QKD protocols. We present a simulation framework to model optical and electrical components for implementing a QKD protocol. The present toolkit aims to model and simulate the optical path of the  DPS-QKD protocol with its imperfections and eventually characterize the optical path. The detailed device-level modeling and analysis capabilities of the present toolkit based on Simulink  and  MATLAB have the potential to provide universal  toolkit for practical design and implementation of generalized QKD protocols compared to earlier works. We report a  novel work on the implementation of a QKD protocol on Simulink  and  MATLAB platform. Further, the absence of any modeling framework for DPS QKD and its simplistic optical schematic made it an obvious choice for the authors. We are hopeful that this work  will pave way for simulating  other QKD protocols from the DPR family.
\end{abstract}

\maketitle

Secure and secret communication techniques are crucial since the advent of information theory in the $20^\textrm{{th}}$ century. Countless people have worked towards building technology for secure communication over the last century. Classical cryptography utilizes the complexity of exponentially hard computational problems to distribute encryption keys.  However, the quantum algorithms pose a severe threat as these problems can be solved in a  polynomial time on a quantum computer. QKD marks a disruptive breakthrough in this journey by exploiting principles of quantum mechanics. It is a technique to generate encryption keys between two parties in a quantum secure way. The first QKD scheme developed in 1984 by Charles Bennett and Gilles Brassard has revolutionized the cryptography industry. Thereafter, several other QKD protocols are proposed and each protocol is subject to rigorous and extensive  security analysis. Many improvements are proposed onto original protocols to enhance the security and range of communication. A common underlying principle among various prepare-and-measure QKD schemes is that the bit information is encoded on a non-orthogonal quantum state. The indistinguishability of non-orthogonal states guarantees the security of the communication. No eavesdropper can overwhelm the quantum principles paving way for robust security of QKD protocols. \\

Modeling and simulation are often used as an efficient means to understand complex systems and their dynamics. By systematically defining and decomposing the complex behavior of interest, one can construct representative models with varying abstraction levels.\\

Initial research focused on the analysis and performance of QKD protocols with idealistic assumptions and limited optical components \cite{zhao2008event, coles2016numerical}. As the technology matured, modeling the practical components and processes involved in entire QKD protocol gained attention \cite{Chatterjee2019, Fan-Yuan2020, Morris2014, Mailloux2015, Mailloux2016, mailloux2015modeling, Buhari2012}. Recently, significant development is made on modeling and simulation of QKD networks with varying topology \cite{Wu2020, Mehic2017}. \\

In 2015, Mailloux \textit{et al.} \cite{Mailloux2015} reported a complete software package with a modular architecture called qkdX for polarization-based BB84 protocol. Recently, Chatterjee \textit{et al.} \cite{Chatterjee2019} presented qkdSim, a software package for modeling and analyzing generic QKD protocols, yet only B92 protocol is simulated. Further, a recent work \cite{Fan-Yuan2020} have prepared a universal framework for QKD modeling. Although these simulation frameworks are claimed to be generalised to simulate any QKD protocol, no progress has been made in simulating the distributed phase reference (DPR-QKD) and continuous variable (CV-QKD). \\

In the present work, we have designed and implemented a QKD simulator for studying the implementation of differential phase shift (DPS) QKD, which falls under the DPR-QKD protocols. We have studied the feasibility of simulating different building blocks of the optical path of the protocol. The purpose of our work is to find a practical and high-precision platform to  analyze the optical path of the protocol and to explore whether the toolkit can be generalized  for all QKD protocols.

DPS protocol is widely accepted for commercial application by virtue of it's simplistic implementation. The present toolkit is built using Simulink and MATLAB, which are widely used for simulation of optical, electronic and opto-electronic systems. Each component and process involved are custom build and designed to assimilate future technological advances. \\

This paper is organised as follows. In section \ref{sec:DPS}, we briefly explain the DPS QKD protocol and the underlying principles. Section \ref{sec:Sim} discusses in detail the modeling scheme and framework in a generalised fashion. We have also listed the assumptions considered in the work. In section \ref{sec:Components}, we unfold the modeling technique of individual components and processes. In section \ref{sec:Discussion}, we present our implementation of DPS QKD in the simulation toolkit and in section \ref{sec:Capabilities} we present the diverse capabilities of the toolkit for analysing a QKD protocol for research objectives. Lastly, we conclude in section \ref{sec:Conclusion}.

\section{DPS-QKD Protocol}\label{sec:DPS}
\begin{figure}[ht]
    \centering
      \includegraphics[width=0.5\textwidth]{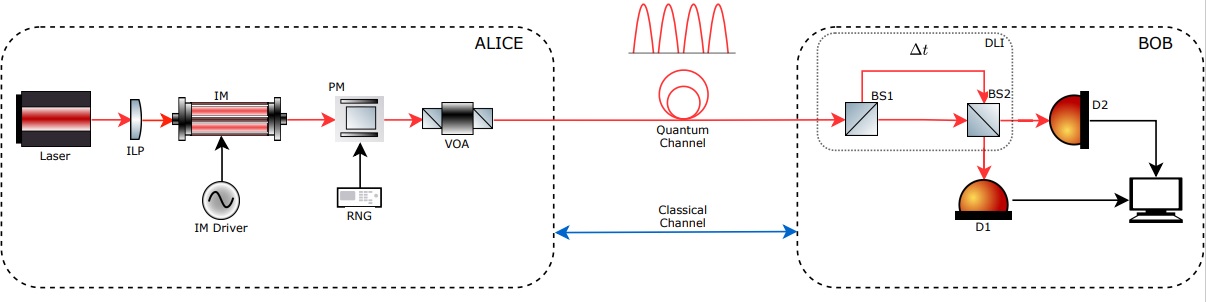}
    \caption{Optical schematic of DPS QKD protocol.}
    \label{fig:DPS}
\end{figure}

Inoue \textit{et al.} introduced DPS QKD in 2002 \cite{Inoue2002, Inoue2003, Inoue2005, Inoue2015}. It is a  prepare-and-measure type of QKD protocol  where, Alice (transmitter) generates  heavily attenuated and phase encoded, train of pulses  and transmits then over a quantum channel to Bob (receiver). The security is based on non-deterministic collapse of a wave function in a quantum measurement. The optical schematic of DPS QKD  is presented in Fig. \ref{fig:DPS}.

The protocol is executed as follows:
\begin{enumerate}
\item Alice generates the state $\ket{\psi}$ which is a train of $N$ coherent pulses  with $\theta$ as the initial phase and $\theta_n$ as encoded phase which takes \{$0,\pi$\}. This quantum state is represented   as:

\begin{equation}
\ket{\psi} = \bigotimes \limits_{n=0}^{N-1} \ket{\alpha e^{i(\theta + \theta_n)}}.
\label{Pulse_train}
\end{equation}

She heavily attenuates the transmission power such that mean photon number per pulse is less than unity.
\item Bob receives the pulse train and passes it through a Delay Line Interferometer (DLI) with a delay which is inversely proportional to repetition rate of pulse train. The interference between adjacent pulses results in selective detection between the two single photon detectors D1(2) for a phase difference of 0($\pi$).

\item Bob tags each click and the corresponding detector. He shares the time information with Alice over prior authenticated classical channel.

\item Considering the time information shared by Bob and the phase encoding information already available with Alice, she can generate a raw data, partially correlated with Bob. This process is the beginning of key reconciliation. Synchronization mismatch with  electronic noise and delay are the major sources of error  causing irregularities in shared data.

\item Bob performs the error estimation procedure with Alice. The discrepancies  are due to imperfections at various levels in the communication system and contribute to the quantum bit error rate (QBER). QBER is defined as the ratio of incorrect bits to received bits. In conservative  security analysis, all the errors are attributed to an eavesdropper. The amount of information exposed  to Eve is estimated using the Shannon’s noiseless coding theorem \cite{welsh1988codes}. For an estimated QBER of $e$, the minimum information exposed is given by

\begin{equation}
   h(e) = -e \log_2 e - (1-e) \log_2(1-e).
\end{equation}

Thereafter, both the parties execute  error-correction to generate correlated bit string on either end.

\item Finally, Alice and Bob perform  privacy amplification to obtain final secure key. The error corrected key is compressed to minimise the information leakage in the raw quantum transmission and during error correction. The extent of compression depends upon the amount of information leak and the eavesdropping strategy.
\end{enumerate}

Numerous enhancements have been discussed \cite{Inoue2015} on the original DPS protocol to improve the transmission range and security against sophisticated eavesdropping attacks. These advancement include variable delay scheme \cite{Sasaki2014} (RR DPS-QKD), implementation  of decoy states \cite{Inoue2008}, a novel  four-level scheme (DQPS QKD)\cite{5224526}, and  entanglement based scheme. Extensive security analysis of these extended schemes is yet an open problem. The present toolkit will be valuable tool for comparative analysis of these modification to original scheme.\\

\section{Simulation Scheme}\label{sec:Sim}
In this section, we provide an overview of the simulation schemes and the assumptions taken to simplify the modeling scheme. We discuss different types of simulation schemes and the problem sets addressed by each criterion and best practices on modeling and simulation. This analysis helps us to classify QKD protocol among the primary simulation classifications and setting the development ground.

\subsection{Overview}
A notable feature of QKD systems is that billions of optical pulses are generated and propagated through the system during the simulation of system operation. While each pulse is most accurately represented as a continuous-time waveform, it is computationally intensive to model a complete QKD system using continuous-time simulation. This forces us to simulate the process in a  discrete-time environment. \\

More broadly, QKD protocols are classified as dynamic, time-invariant, nonlinear processes \cite{Morris2014}. The dynamic nature is due to the dependence of output over input signals. Although the system behavior changes with time, it is invariant to the universal time; thus, it is considered as  time-invariant. Nonlinearity arises from the various complex physical components and processes. A QKD protocol architecture is divided among five discrete parts: (a) source, (b) preparation, (c) transmission, (d) detection, and (e) post-processing. This particular order of events results in a systematic flow of signals in the discrete-event based environment. \\

An essential characteristic of a QKD protocol is its probabilistic nature, where the outcomes of an event can be associated with  probability. QKD is inherently a random process represented in its subsystems like a laser, beam splitter, detector, etc. The probabilistic nature is the foundation of QKD    thus, it is crucial to model the optical path. For each stochastic process, random variables are generated from a (pseudo-random number generator) PRNG to emulate the physical processes \cite{bremaud2012introduction}. \\

A dynamic, time-invariant, nonlinear, stochastic, and discrete-time simulation is the appropriate model characteristic of a QKD protocol. These are the imperative requirement in a simulation framework and is essential to describe before constructing the simulation model. Simulink is one such package that accommodates all these features quickly and has an open structure to design and implement a new system. Simulink is best suited for the process with its vast simulation toolkits on communication and signal processing. \\

Independent modules based upon model characteristics are developed corresponding to the source, detection, and transmission components used in the actual setup. The modules are interconnected for the flow of logic and mimic the photons' path in the actual experimental setup. The simulation aims to illustrate the QKD process alongside various non-idealities due to device imperfections. Another critical objective is to simulate hacking attacks on the system, helping us perform a security analysis of practical QKD protocols. \\

Each module is composed of sub-modules corresponding to physical components and processes. Each sub-module is parameterized by certain physical variables affecting the dynamics. They are categorized as either user inputs or set parameters. User inputs refer to the user's individual choices, whereas the set parameters refer to the specification of the various components.

\subsection{Assumptions}
The simulation model creates a virtual experimental implementation of DPS QKD by modeling imperfections, but it is not exhaustive. We have to make certain assumptions to simplify the system. We have listed below some of them:

\begin{itemize}
\item Optical source is assumed to produce monochromatic coherent light with appropriate spectral linewidth.

\item Optical transmission medium is a single mode fibre and the signal propagates only in the allowed fundamental $\textrm{LP}_{01}$ mode.

\item The optical signal is represented by an electric field propagating in z direction and it's components in two polarisation axis (x and y) where, x being the faster and y is the slower polarisation axis.

\item Interference of optical signal occurs only at DLI, and is ignored elsewhere.

\item Optical loss in between the two components is considered negligible, however, insertion loss of individual components is included. Also, all the optical components are assumed to be perfectly aligned.

\item Both Alice and Bob subsystem  runs on a single clock, which is  the simulation clock. This eliminates the clock synchronisation imperfections.

\item Presently, no eavesdropping strategy or attack is considered for the simulation but this would be included in future works.

\item Sifted keys can be obtained from the simulator, however, the  error correction and privacy amplification algorithms can be applied separately.

\end{itemize}
\vspace{1em}

\section{QKD unit modules} \label{sec:Components}
In this section, we have described the modeling scheme for each physical component. We have also tabulated the variables contributing to the modeling scheme of each component alongside.

\subsubsection{Laser}
Laser is a complex self-consistent system that is capable of demonstrating a wide range of dynamics. Semiconductor lasers are most frequently used optical source in optical fibre communication technologies due to their compact size, low power consumption and affordability. Distributed feedback (DFB) laser are prominently used, credit to their superior performances, narrow spectral width and low noise. \\

The operating dynamics of a DFB semiconductor laser is defined by a set of rate equations for the interaction of photons and charge carriers inside the active cavity region \cite{DFBLaser}. They are derived from Maxwell’s equations using quantum-mechanical approach for the induced polarization. These single mode optical laser rate equations (\ref{carrier_density}, \ref{photon_density}) are used for the simulation of the frequency chirp and output power waveform. For a single-mode laser, the rate equations are given below:

\begin{equation}
\frac{dN(t)}{dt} = \frac{I(t)}{eV_a} - \frac{N(t)}{\tau_n} - g_0 \frac{N(t) - N_0}{1 + \epsilon_C S(t)} S(t)
\label{carrier_density}
\end{equation}

\begin{equation}
\frac{dS(t)}{dt} = \left(\Gamma g_0 \frac{N(t) - N_0}{1 + \epsilon_C S(t)} - \frac{1}{\tau_p} \right) S(t) + \frac{\beta \Gamma N(t)}{\tau_n}
\label{photon_density}
\end{equation}

\begin{equation}
P(t) = \frac{S(t)V_a\eta_0h\nu}{2\Gamma\tau_p}
\label{laser_power}
\end{equation}

Equ. (\ref{carrier_density}, \ref{photon_density}) are coupled non-linear differential equations between the charge carrier density, N(t) and photon density, S(t). The carrier density N(t) increases due to the injection current I(t) into the active layer volume $V_a$ and decreases due to stimulated and spontaneous emission of photon density S(t). Similarly, the photon density S(t) is increased by stimulated and spontaneous emission S(t) and decreased by internal and mirror losses. The time variations of the output optical power are related to the photon density as shown in equ. (\ref{laser_power}). The resultant electric field is calculated in equ. \ref{elec} where, the amplitude $A(t) = \sqrt{\frac{2P(t)}{\epsilon V_a}}$.

\begin{equation}
E(t) = A(t) e^{i(\omega_c t + \phi(t))}
\label{elec}
\end{equation}

\begin{table}[ht]
\begin{tabular}{p{0.1\textwidth} p{0.3\textwidth}}
\toprule
\textbf{Symbol} & \textbf{Parameter} \\
\midrule
$\Gamma$ & Optical confinement factor \\
$g_0$ & Gain coefficient \\
$N_0$ & Carrier density at transparency \\
$\epsilon_C$ & Gain compression factor \\
$\tau_p$ & Photon lifetime \\
$\beta$ & Spontaneous emission coupling factor \\
$\tau_n$ & Electronic carrier lifetime \\
$e$ & Electronic charge \\
$V_a$ & Active region volume \\
$\eta_{\textrm{DFB}}$ & Differential quantum efficiency \\
$N_0$ & Equilibrium carrier density \\
$h$ & Plank's constant  \\
$c$ & Speed of light \\
$\lambda_0$ & Central wavelength \\
$\epsilon$ & Permittivity of active region \\
$I(t)$ & Injection current  \\
$\sigma_I$ & Injection current standard deviation \\
$ \Delta \omega $ & Spectral linewidth \\
$\textrm{RIN}$ & Relative intensity noise \\
$\delta_{\lambda}$ & Temperature variation \\
$\Delta f$ & System bandwidth \\
$T_L$ & Laser set temperature \\
$\phi_0$ & Initial absolute phase \\
\bottomrule
\end{tabular}
\end{table}

A laser exhibit two types of noise, relative intensity noise (RIN) ($\delta P(t)$) and  phase noise ($\delta \phi (t)$). In our implementation, RIN value is obtained from product data sheet and is assumed to be white gaussian. The phase noise is related to the spectral line width and inverse of clock frequency. Gaussian noise of calculated variance is added to the power and the phase to replicate actual system imperfections. The rate equation model along-with emulation of laser noise and temperature dependence demonstrates response of a monochromatic laser with a reasonable fidelity. The laser sub-module outputs electric field signal ($E_x (t)$ and $E_y (t)$) as a bus signal. In QKD protocols based on phase encoding the linewidth plays a critical role. In DPS QKD, improper selection of linewidth impacts the system QBER and its stability \cite{HONJO20115856}. It is reported that linewidth should be less than $ 0.35\% $ of the free-spectral-range (FSR) of DLI to achieve QBER less than $0.5\%$. The toolkit provides the flexibility to model and study the same leading to improvement of optical path performance and minimizing the system QBER.

\subsubsection{In-line Polariser}
An in-line polariser pass linearly polarized light while blocking the orthogonal polarization from an unpolarised (or randomly polarized) light source. The orthogonally polarized light is attenuated with a desired  extinction ratio such that only a single principle polarisation mode traverse through the fibre.

\begin{table}[ht]
\begin{tabular}{p{0.1\textwidth} p{0.3\textwidth}}
\toprule
\textbf{Symbol} & \textbf{Parameter} \\
\midrule
$\textrm{ILP}_{\textrm{loss}}$ & Insertion loss \\
$\textrm{PER}$ & Polarisation extinction ratio \\
\bottomrule
\end{tabular}
\end{table}

\subsubsection{Intensity Modulator (IM)}
\begin{figure}
    \centering
    \includegraphics[width =0.5\textwidth]{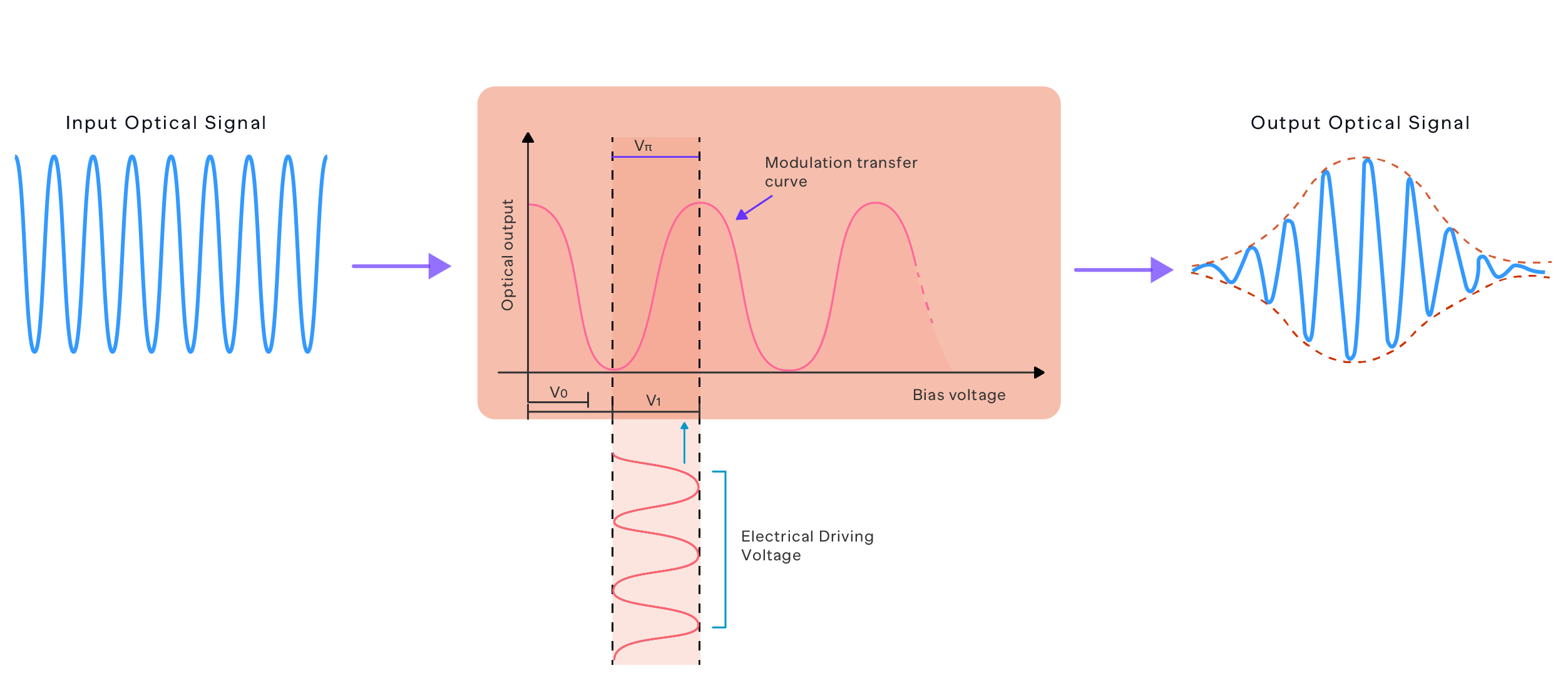}
    \caption{Transfer function of IM.}
    \label{fig:IM_TF}
\end{figure}

IM is used to externally modulate the optical signal from source to result in a pulsed output. A mach-zehnder interferometer structure is used to modulate the optical signal, both the arms experience a phase shift and are recombined. The two electrodes are biased with a RF signal causing a phase shift of $\phi_1(t)$ and $\phi_2(t)$ for the two branches. The output field of the light wave carrier is represented by
\begin{equation}
E_o(t) = \frac{1}{2} E_i \left( e^{i\phi_1(t)} + e^{i\phi_2(t)} \right).
\label{IM_equ}
\end{equation}
For simplicity, equal and opposite potential (V(t)) is applied to the two electrodes, thus the transfer function is reduced to $ E_o(t) = E_i \textrm{cos}(\frac{\pi V(t)}{V_{\pi}})$. $V_\pi$ is a device parameter for $\pi$ phase shift. Transfer function is illustrated in Fig. \ref{fig:IM_TF}.

$V_1$ and $V_0$ in Fig. (\ref{fig:IM_TF}) represent the voltages corresponding to max and min of the optical pulse as shown in Fig.  (\ref{fig:IM_TF}). By adjusting these bounds, a desired extinction ratio
can be achieved. \\

The spectrum of the modulated signals includes sets of symmetric side-bands arranged around the laser carrier frequency $f_o$. The side-bands are displaced from the laser carrier peak frequency at integer multiples of the modulation frequency $f_m$ $(f_o \pm n f_m$ with $n = 1, 2, ...).$ The relative heights of the side-bands are a function of the modulation depth, which is in turn a function of the peak-to-peak value of the RF driving voltage. As a result of these side-bands a multi-model interference appears at the DLI on Bob's side (discussed in section \ref{sec:Discussion}).

\begin{table}[ht]
\begin{tabular}{p{0.1\textwidth} p{0.3\textwidth}}
\toprule
\textbf{Symbol} & \textbf{Parameter} \\
\midrule
$\textrm{IM}_{\textrm{loss}}$ & Insertion loss \\
$\textrm{IM}_{\textrm{ER}}$ & Desired extinction ratio \\
$\textrm{FWHM}$ & Full width half maximum \\
$V_{(\pi)\textrm{IM}}$ & Bias voltage for $\pi$ phase shift \\
$V_{\textrm{DC}}$ & DC bias input to IM \\
$V_{\textrm{RF}} (t)$ & RF input to IM \\
\bottomrule
\end{tabular}
\end{table}

\subsubsection{Phase Modulator (PM)}
PM modulates the phase of optical carrier signals by passing it through an  opto-electrical wave-guide under a RF voltage. As the refractive index of the opto-electrical medium changes with voltage applied, any desired phase shift can be provided to the input signal. An important parameter to characterise PM is $V_\pi$, the RF bias voltage required to give a phase shift of $\pi$. The resultant electric field is given by, $ E_o(t) =  E_i e^{i \frac{\pi V(t)}{V_{\pi}}}$.

\begin{table}[ht]
\begin{tabular}{p{0.1\textwidth} p{0.3\textwidth}}
\toprule
\textbf{Symbol} & \textbf{Parameter} \\
\midrule
$V_{(\pi)\textrm{PM}}$ & Bias voltage for $\pi$ phase shift \\
$V_{\textrm{bias}}$ & Drift in bias voltage \\
$PM_{\textrm{loss}}$ & Insertion loss \\
\bottomrule
\end{tabular}
\end{table}

\subsubsection{Variable Optical Attenuator (VOA)}
VOA attenuates the optical signal to quantum level, reducing the mean photon number to less than unity. The attenuator provide finer control over the quantum amplitude. This is a critical parameter and its precise tuning is very essential from security of quantum channel.

\subsubsection{Optical Fiber}
Light propagating through an optical fiber mainly undergoes both attenuation and distortion of signal. Optical fiber has a finite number of guided propagation modes, with defined spatial structures. V-number is a dimensionless parameter widely used to determine the fraction of the optical power in a certain propagation mode. V-value below 2.405 indicates a single mode propagation. For a 1550 nm telecommunication wavelength and SMF-28 fiber, V-number $= 2.325$ suggesting a single mode propagation. We assume light propagates only in LP01 mode with transverse profile approximated as Gaussian. \\

Transmission of an optical pulses along a single mode optical fiber (SMF) is governed by the Non-Linear Schrodinger equation (NLSE), derived from Maxwell equations \cite{Binh2014}. The influence of optical fiber can be classified into (a) linear effects, which are wavelength depended and  (b) non-linear effects, which are intensity (power) depended. \\

Major impairments of optical signals transmitted via single mode fiber are mainly caused by linear effects - dispersion and attenuation. Attenuation limits power of optical signals and represents transmission losses. Dispersion causes spreading of optical pulses in time domain and phase shifting of signals at the fibre end. Based upon the cause of dispersion it is classified as:

\begin{itemize}

\item Chromatic dispersion (CD): It is the dependence of group velocity associated with the fundamental mode on the frequency of signal. It is also known as group velocity dispersion (GVD). It can be further distinguished as either material dispersion or wave-guide dispersion.

\item Higher-order dispersion: The non zero higher derivatives of the total dispersion curve causes pulse broadening at zero-dispersion wavelength.

\item Polarization mode dispersion (PMD): The birefringence of optical fiber between two principle polarisation axis results in signal delay.
\end{itemize}

Nonlinear effects are crucial for long haul optical signal transmission. The intensity dependence phenomenon of fiber refractive index is known as the Kerr effect and is the cause of fiber nonlinear effects \cite{Binh2014}. The power dependence of refractive index is expressed as:

\begin{equation}
    \eta ' = \eta +  \gamma \frac{P}{A_{\textrm{eff}}}
\end{equation}
where, P is the average optical power of the guided mode, $\gamma$ is the fiber nonlinear coefficient and $A{\textrm{eff}}$ is the effective area of the fiber. Kerr nonlinearity effects include:

\begin{itemize}
\item Self-phase modulation (SPM): Optical intensity causes a nonlinear a time-dependent phase shift and pulse acquires a chirp.

\item Cross-phase effect (XPM): Non-linear phase shift similar to SPM but caused by the interaction with another beams present in the fiber.

\item Four-wave mixing (FWM): Due to third order nonlinearity, mixing of different frequency components propagating together generates additional frequency components.
\end{itemize}

Apart from this, nonlinearity induces inelastic scattering of incident photons into lower energy photons, causing a downward shift of frequency. Such effects found in a optical fiber are - stimulated Brillouin scattering (SBS) and stimulated Raman scattering (SRS). Effects of SRS and SBS are only noticeable with a high optical power and thus, can be  ignored in our application. On the other hand, FWM and XPM becomes negligible in SMF due to high local dispersion. Thus, SPM is usually the only dominant nonlinear effect at low transmission power. Since, operating power in QKD is significantly low, it is safe to neglect all other effects. \\

The numerical method used to solve the NLSE is known as the split-step Fourier method (SSFM) \cite{Binh2014}. It accurately models both linearity and non-linearity of a SMF. In SSFM, the fiber length is divided into small segments $dz$, both linear and non-linear effects within the small length $dz$ are small and thus considered mutually independent of each other. So, the NLSE is transformed into operators formulated as follows:

\begin{equation}
\frac{dA}{dz} = A(\hat{D} + \hat{N})
\end{equation}
\begin{equation}
\hat{D} = - \frac{\alpha}{2} + \frac{i}{2} \beta_2 \frac{\partial ^2 }{\partial t^2} + \frac{1 }{6} \beta_3 \frac{\partial ^3 }{\partial t^3} \pm \frac{i \Delta \tau}{2} \frac{\partial }{\partial t}
\end{equation}
\begin{equation}
\hat{N} = i \abs{A \sqrt{\gamma}}^2
\end{equation}

where, $\hat{D}$ represents the linear differential operator accounting for absorption ($\alpha$), CD ($\beta_2$), higher order dispersion ($\beta_3$) and PMD ($\Delta \tau$). $\hat{N}$ is the non-linear operator representing SPM. The accuracy of SSFM can be improved by sandwiching the effect of fiber non-linearity by dispersion effect in each segment $dz$. \\

\begin{table}[ht]
\begin{tabular}{p{0.1\textwidth} p{0.3\textwidth}}
\toprule
\textbf{Symbol} & \textbf{Parameter} \\
\midrule
$L$ & Fibre length \\
$\alpha$ & Attenuation per km  \\
$\beta_2$ & $2^{\textrm{nd}}$ order CD factor \\
$\beta_3$ & $3^{\textrm{rd}}$ order CD factor \\
$\Delta \tau$ & Differential group delay \\
$\gamma$ & Non-linear coefficient \\
NFFT & Number of samples for FFT \\
$dz$ & Step size for SSFM \\
\bottomrule
\end{tabular}
\end{table}

\subsubsection{Delay Line Interferometer (DLI)}
\begin{figure}[ht]
\centering
\includegraphics[scale=0.3]{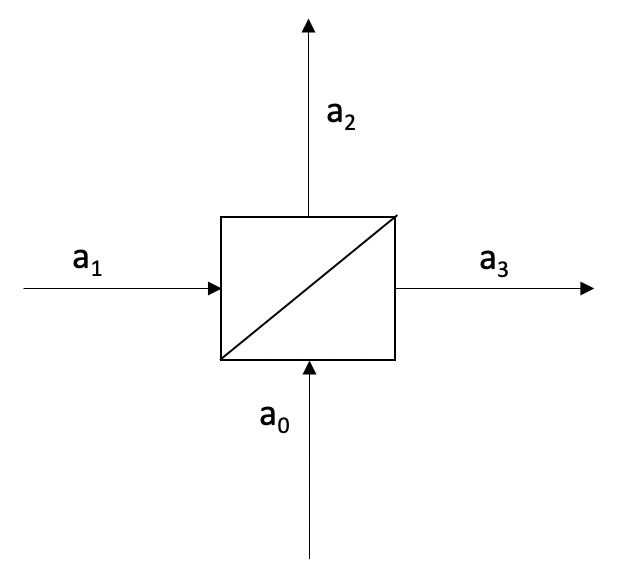}
\caption{Input and output ports of a beam splitter.}
\label{fig:BS}
\end{figure}

DLI,  encompasses interference of adjacent pulses to demodulate the information stored in relative phase  of pulses. A typical DLI comprises of two 50:50 beam splitter (BS) each with two inputs and two outputs as shown in Fig. \ref{fig:BS} (BS1 assumes vacuum state as second input). The path-splitting operations of BS are parameterized by reflectivity (r) and transmissivity (t) which follows $\abs{r}^2 + \abs{t}^2 = 1$.

\begin{gather}
\begin{pmatrix} \hat{a}_0^{\dagger} \\ \hat{a}_1^{\dagger} \end{pmatrix} \rightarrow
\begin{pmatrix}
t & ir \\
ir & t
\end{pmatrix}
\begin{pmatrix} \hat{a}_2^{\dagger} \\ \hat{a}_3^{\dagger} \end{pmatrix}
\label{BS1}
\end{gather}

\begin{gather}
\begin{pmatrix} \hat{a}_0 \\ \hat{a}_1 \end{pmatrix} \rightarrow
\begin{pmatrix}
t & -ir \\
-ir & t
\end{pmatrix}
\begin{pmatrix} \hat{a}_2 \\ \hat{a}_3 \end{pmatrix}
\label{BS2}
\end{gather}

Equ. (\ref{BS1}) and (\ref{BS2}) provide the correlation between the corresponding state creation and annihilation operators. Employing above equations with displacement operator, the input and output states are given by:
\begin{equation}
\ket{\textrm{in}} = \ket{\alpha}_1 \ket{\beta}_0
\end{equation}

\begin{equation}
\ket{\textrm{ou}} = e^{i\phi} \ket{t\beta + ir\alpha}_2 \ket{t\alpha + ir\beta}_3
\label{interference}
\end{equation}

Equ. (\ref{interference}) represents the interference of the adjacent pulses in an ideal case. However, the optical signal is comprised of multiple side bands due to the external modulation. Thus, a multi-modal interference occurs at the second beam splitter and the resultant power is given by:
\begin{equation}
P(t) = \abs{\sum_{i}{E^{(i)}_x(t)} +  \sum_{j}{ E^{(j)}_y(t)}}^2
\end{equation}

where, $E^{(i)}_x(t)$ and $E^{(i)}_y(t)$ are $i^{th}$ mode of electric field in x and y polarisation field respectively. The central frequency is represented by $E^{(0)}_x(t)$ and $E^{(0)}_y(t)$. This imperfect interference at the DLI causes a reduction in the output visibility defined by

\begin{equation}
V = \frac{I_{\textrm{max}} - I_{\textrm{min}}}{I_{\textrm{max}} + I_{\textrm{min}}}
\end{equation}

\begin{table}[ht]
\begin{tabular}{p{0.1\textwidth} p{0.3\textwidth}}
\toprule
\textbf{Symbol} & \textbf{Parameter} \\
\midrule
$\Delta t$ & Time delay \\
t & Transmissivity \\
r & Reflectivity \\
$\textrm{DLI}_{\textrm{loss}}$ & Insertion loss \\
\bottomrule
\end{tabular}
\end{table}

\begin{figure*}[t!]
    \centering
    \includegraphics[width = \textwidth]{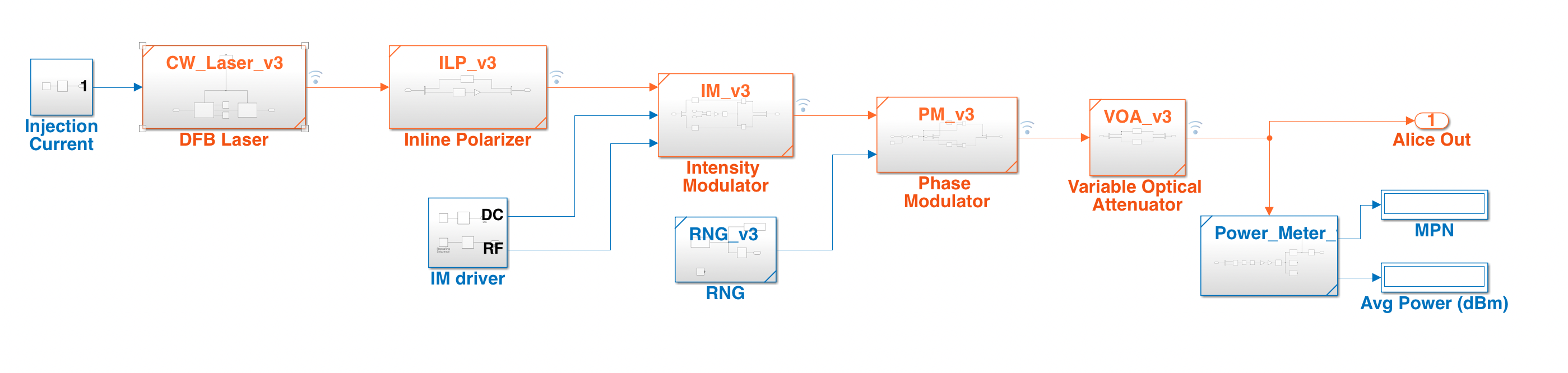}
    \caption{Block diagram from Simulink for Alice's subsystem in DPS QKD}
    \label{fig:alice}
\end{figure*}
The visibility plays a critical role in system QBER and system stability for practically, all phase encoding based QKD protocols. The QBER is given by
\begin{equation}
\frac{1-V}{2}
\end{equation} which implies that if visibility is $90\%$ then optical QBER is $5\%$ . The impact of proper implementation of DLI whether based on PLC or free space DLI unit can be studied by the simulating the DLI unit module accordingly.

\subsubsection{Single Photon Detector (SPD)}
For a prepare-and-measure QKD protocol, measuring the quantum state is the most decisive step. In DPS, Bob's setup comprises of two single-photon detectors (SPD) placed at the two outputs of DLI. Since, the information is encoded in the pulse train with mean photon number less than unity, we need single photon detectors for measuring the qubit send by Alice. single photon avalanche detector (SPAD) are mostly used for this application due to it's relative simplicity and affordability as compared to superior superconducting nano-wire single photon detectors (SNSPD). \\

In this work, we try to model and simulate the SPAD incorporating it's complex behavior depending on deadtime, detection efficiency, dark count rate, after-pulse probability, temperature and time jitter. Many different approaches have been made to accurately model a SPD \cite{Liu2012, Kang2003, Rohde2006, Cheng2016}. We have used a probabilistic approach to model the SPD running in continuous mode. Firstly, the mean photon number in each pulse is calculated using the energy content of each pulse. The detection inefficiency of SPD is equivalent to the transmission losses. Thus, the MPN arriving at detector equals to $\eta \mu$ and detector detects all these photons. We also add background photon ($N_b$) to the incoming photons. The Poisson distribution of photon given by Equ. (\ref{poisson}) is used to calculate the probability of more then one photons in the pulse, $P_p = 1 - e^{\eta \mu + N_b}$.

\begin{equation}
\ket{\alpha} = e^{-(\eta \mu + N_b )/2} \sum_{n} \frac{(\eta \mu + N_b)^{n/2}}{\sqrt{n!}} \ket{n}
\label{poisson}
\end{equation}

A detection event in SPAD is characterized  by multiple phenomena such as dark count, after pulsing, deadtime etc. To model the after-pulsing effect as a function of number of interval pulses $n$, we can use an exponential distribution function given by:

\begin{equation}
P_{\textrm{ap}}(n) = p_0 e^{-\textrm{an}}
\end{equation}

The two parameters  $p_0$ and $a$ are set to 0.0317 and 0.00115  \cite{Liu2012}, respectively. Given the dark count probability ($P_d$) from detector specification sheet the probability of registering a detection event, $P_{\textrm{click}}$ can be calculated as:

\begin{multline}
P_{\textrm{click}} = P_p + P_{\textrm{ap}} + P_d - P_p P_{\textrm{ap}} \\ - P_{\textrm{ap}} P_d
- P_d P_p + P_p P_{\textrm{ap}} P_d.
\end{multline}

To replicate the probabilistic nature of photon detection, a random number is generated between 0 and 1. If the random number is less then $P_{\textrm{click}}$, a click is registered. Once, a photon click is registered the detector goes OFF for a certain time-interval, known as detector dead time ($\tau$) and the $P_{\textrm{click}}$ is set to zero for the interval. After each iteration the after-pulse probability is updated. The click interrupt signal from detector is added with a suitable time jitter, similar to realistic devices.

\begin{table}[ht]
\begin{tabular}{p{0.1\textwidth} p{0.3\textwidth}}
\toprule
\textbf{Symbol} & \textbf{Parameter}\\
\midrule
$N_b$ & Background photons per pulse \\
$\tau$ & Deadtime \\
$P_d$ & Dark count probability \\
$p_0$ & After-pulse probability pre factor \\
$a$ & After-pulse probability exponential factor \\
$\eta$ & Detector efficiency \\
$SPD_{Jit}$ & Time jitter \\
\bottomrule
\end{tabular}
\end{table}

\section{DPS-QKD Simulation} \label{sec:Discussion}
In this section, we present the modeling architecture of DPS QKD protocol. We have considered the original DPS scheme \cite{Inoue2005} already described in section \ref{sec:DPS} with standard single-mode telecommunication fiber  as medium of communication. This exercise provides a distinct overview of simulation toolkit in evaluating QKD system architecture. Similar to a QKD protocol, the communication system is divided into Alice and Bob, comprising of optical components discussed in section \ref{sec:Components}. Each component sub-module is a physical representation of components and together form the quantum communication path.

\subsection*{Alice}
Alice subsystem in DPS prepares a quantum state with secret bit information encoded in phase. Fig.  \ref{fig:alice} represents the Alice subsystem's modeling architecture. A constant injection current to DFB laser module generates a continuous optical signal as a photon source. Injection current can be altered to replicate a direct modulation scheme also. Inline polariser suppresses an orthogonal polarisation mode and results in a linearly polarised signal. This reduces the dispersion due to polarisation while transmission. \\

\begin{figure}[ht]
    \centering
    \includegraphics[width = 0.5\textwidth]{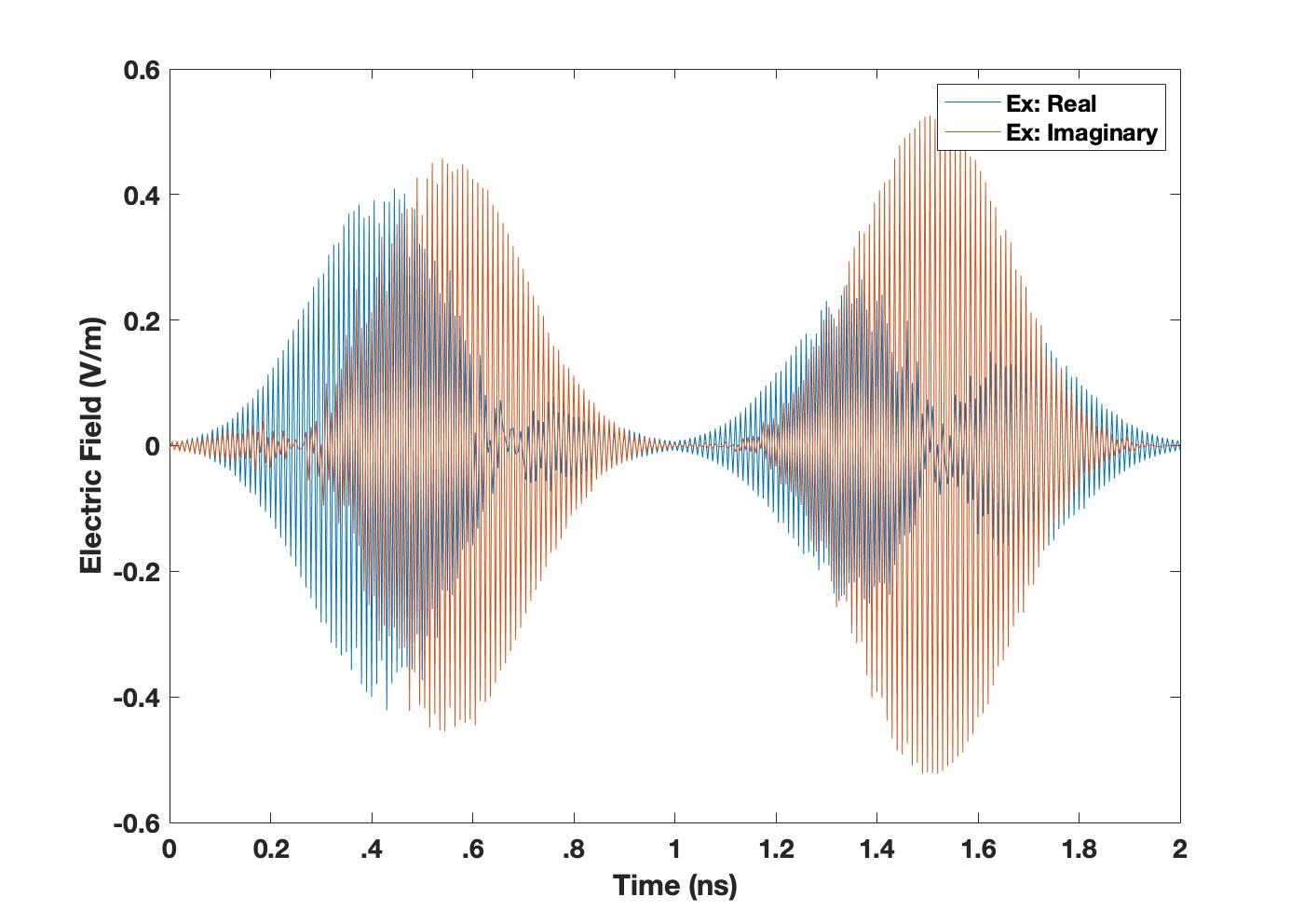}
    \caption{Modulated electric field signal from IM illustrating photon wave packets.}
    \label{fig:IM_out}
\end{figure}

At the next step, IM transforms the continuous signal into a pulse train (Fig. \ref{fig:IM_out}) where, individual pulse represent a coherent quantum state. IM is driven by $V_{DC}$ and $V_{RF}$, governing the resultant extinction ratio and pulse profile. Further, PM encodes random binary bit information into each coherent pulse precisely, 0(1) bit from the Simulink's Bernoulli random number generates translates into a phase shift of 0 ($\pi$). Finally, the power level is reduced to quantum level using the VOA. For regular monitoring of the average power and mean photon number we have designed a power meter with live display of physical quantities.

\subsection*{Bob}
\begin{figure}[ht]
    \centering
    \includegraphics[width = 0.5\textwidth]{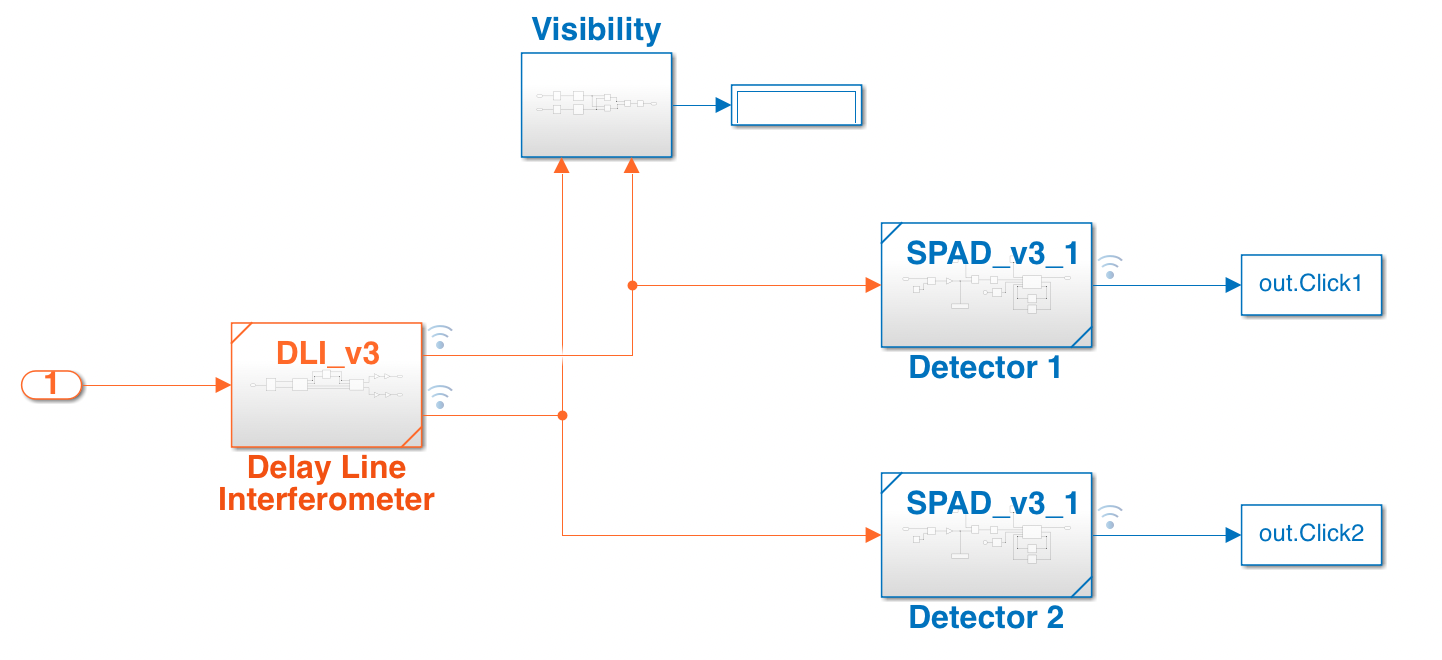}
    \caption{Block diagram from Simulink for Bob's subsystem in DPS QKD}
    \label{fig:bob}
\end{figure}

\begin{figure}[ht]
    \centering
    \includegraphics[width = 0.5\textwidth]{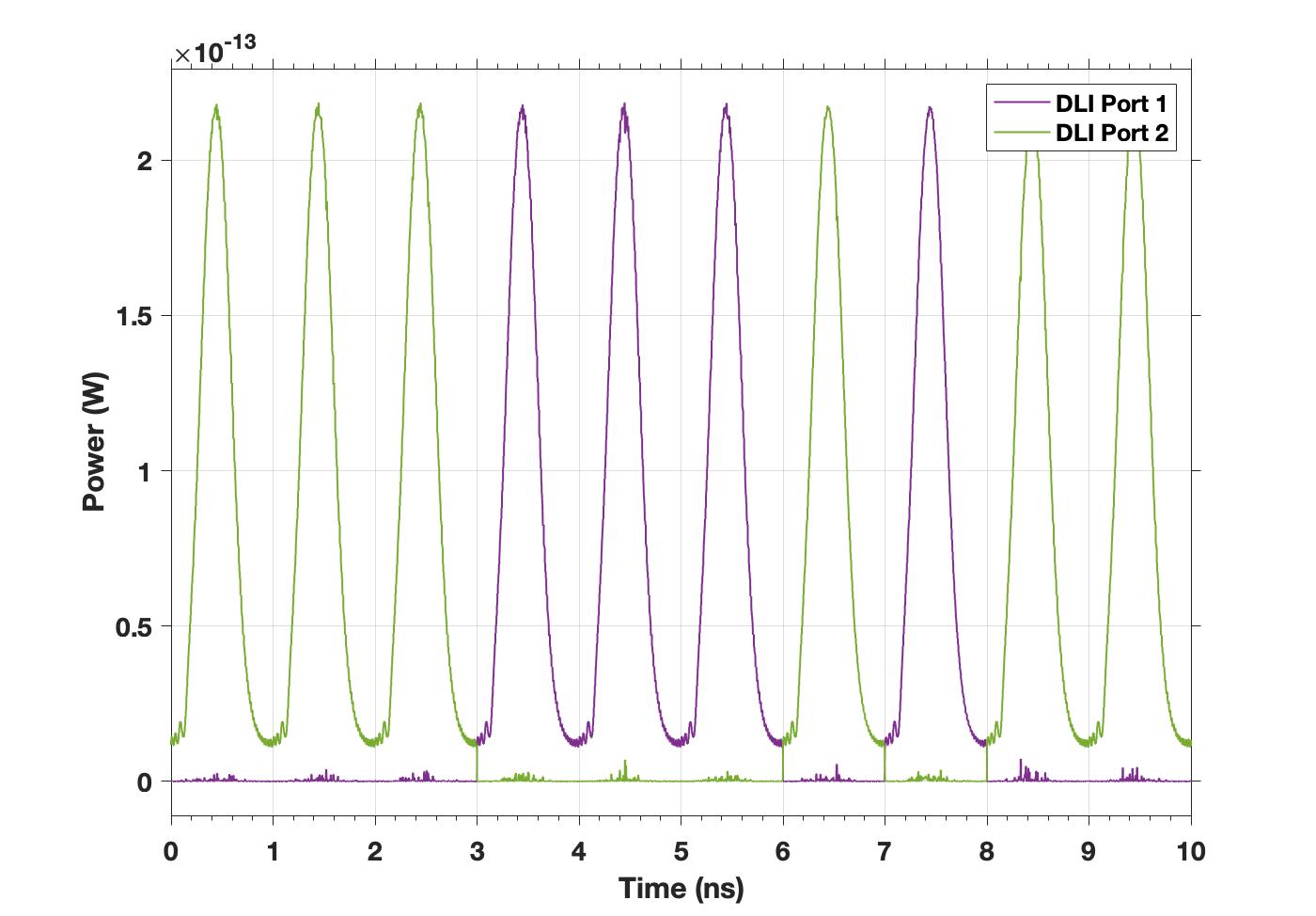}
    \caption{Optical signal at the two output ports of DLI.}
    \label{fig:DLI_out}
\end{figure}

In Fig. \ref{fig:bob} we have presented the Bob's subsystem, it receives the weak coherent pulse train  transmitted by Alice over an optical fibre. To  extract the encoded bit information he passes it through a DLI. The  demodulation of the signal is achieved  by interference of adjacent pulses. In Fig. \ref{fig:DLI_out}  we have shown the output waveforms from two ports of DLI as observed from the simulator. Eventually DLI  bifurcates the signal based on relative phase difference. The optical noise like poor extinction ratio at DLI, improper phase encoding  and non-idealities during signal propagation results in a poor interference and can be quantified by the visibility described earlier. SPDs at two ports of DLI detects single photons based upon the probabilistic model.

\section{Capabilities} \label{sec:Capabilities}
The simulation toolkit aims to better understand the fallout of device imperfection on a QKD protocol. The performance of a QKD protocol is generally gauged by parameters for example QBER, secure key rate and key asymmetry. These parameters are calculated after each cycle of communication between Alice and Bob. QBER is the most essential parameter, defined as the ratio of false bit and total shared bits. Factors contributing to QBER include laser line width, extinction ration of IM, fiber losses, dark counts, optical visibility, phase jitter and phase modulation demodulation errors. By carefully controlling the imperfections, simulation is used to study and analyze the underlying relations for the performance parameters.

\begin{figure}
    \centering
    \includegraphics[width = 0.5\textwidth]{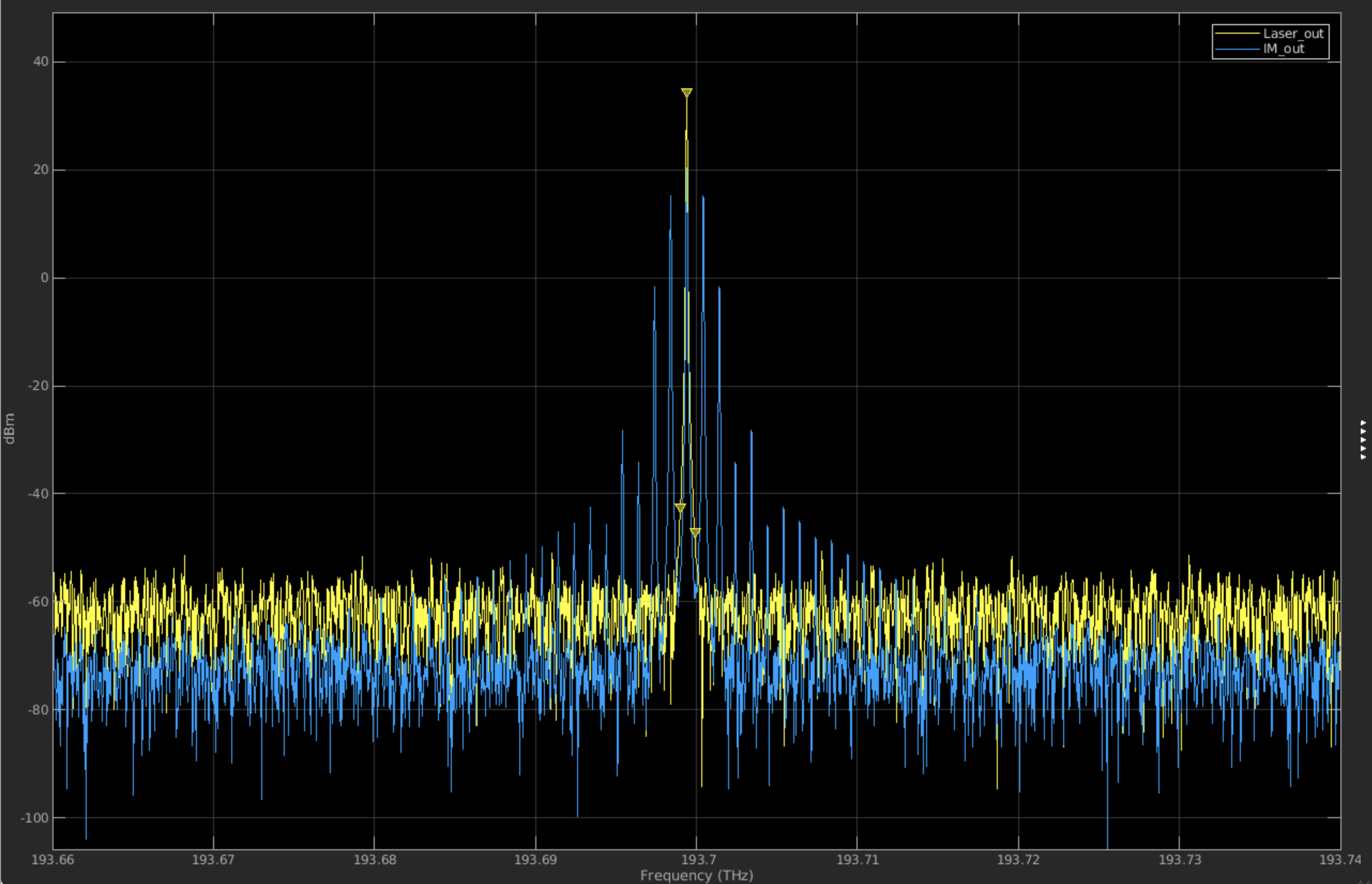}
    \caption{Spectrum analysis of optical signal from Laser (yellow) and IM (blue) showing side-bands in optical signal.}
    \label{fig:spectrum}
\end{figure}

Simulation also provide visualisation tools for optical signal at various step. Parameter which are difficult to calculate theoretically such as signal to noise ratio, visibility, extinction ratio and polarisation extinction can be estimated from the toolkit for enhanced implementations. \\

Theoretical studies on QKD often miss studying the frequency spectrum of the transmitted signal. With several wavelength dependent eavesdropping attacks \cite{Li2011}, it is crucial to analyse the spectrum of optical communication. We have used Simulink's spectrum analysis tool \cite{spectrum} to examine the intricate details of optical spectrum.\\ Although, the optical source is said to produce single mode optical signal, but due to external modulation and scattering, side bands are generated along with the carrier frequency as shown in Fig.  \ref{fig:spectrum}. The spectrum analyser incorporated in the simulation toolkit targets this problem and can be a useful tool for research activities.

\section{Conclusion} \label{sec:Conclusion}
In this work, we  present the framework for MATLAB based DPS QKD simulation toolkit, with a bottom-up approach for modeling.  This comprehensive approach of modeling optical and electrical components provides the flexibility to push it as a universal toolkit for all discrete variable based prepare-and-measure and entanglement based QKD protocols. The model helps us in studying the impact of different characteristics of components and its impact on system QBER and system stability. The present toolkit based on Simulink  and  MATLAB offers a user friendly tool leveraging MATALB's extensive development and analysis tools to accurately model the physical devices and their shortcomings. We report for the first time modeling of physical components from the first principles. We note that earlier works \cite{Chatterjee2019, Mailloux2015} posses limited capabilities to analyze the optical path from source generation to signal detection, particularly, analyzing the effects on signal propagation at intermediary  points.  \\

Future work involves  modeling an extensive library of optical components and perform simulation of other QKD protocols based on particularly, prepare-and-measure method. We also intend to simulate quantum attacks on QKD system. This  will help us  map the   loopholes of the system with quantum attacks. Thus,  it will create   better understanding of  Eve's strategy in real-life and effectiveness of proposed countermeasures. Our objective is to build a universal standalone QKD simulation toolkit to help enhance designing and implementing for better performance of QKD protocols in practice.

\bibliography{qkd_sim} 

\end{document}